\documentclass[letter,
               ]{jacow}
\usepackage{pdfpages,multirow,ragged2e} %
\makeatletter%
	\ifboolexpr{bool{xetex}}
	 {\renewcommand{\Gin@extensions}{.pdf,%
	                    .png,.jpg,.bmp,.pict,.tif,.psd,.mac,.sga,.tga,.gif,%
	                    .eps,.ps,%
	                    }}{}
\makeatother

\ifboolexpr{bool{xetex} or bool{luatex}} 
 {}                                      
 {\usepackage[utf8]{inputenc}}           

\usepackage[USenglish]{babel}
\usepackage{amsmath}
\ifboolexpr{bool{jacowbiblatex}}%
 {%
  \addbibresource{jacow-test.bib}
  \addbibresource{biblatex-examples.bib}
 }{}
\begin{document}

\title{Direct RF Sampling based LLRF Control System for C-band Linear Accelerator}

\author{C. Liu\thanks{chaoliu@slac.stanford.edu}, R. Herbst, B. Hong, L. Ruckman, E. A. Nanni \\ SLAC National Accelerator Laboratory, Menlo Park, California, USA \\
		}
	
\maketitle

\begin{abstract}
   Low Level RF (LLRF) control systems of linear accelerators (LINACs) are typically implemented with heterodyne based architectures, which have complex analog RF mixers for up and down conversion. The Gen 3 Radio Frequency System-on-Chip (RFSoC) device from AMD Xilinx integrates data converters with maximum RF frequency of 6~GHz. This enables direct RF sampling of C-band LLRF signal typically operated at 5.712 GHz without any analogue mixers, which can significantly simplify the system architecture. The data converters sample RF signals in higher order Nyquist zones and then up or down convert digitally by the integrated data path in RFSoC. The closed-loop feedback control firmware implemented in FPGA integrated in RFSoC can process the base-band signal from the ADC data path and calculate the updated phase and amplitude to be up-mixed by the DAC data path. We have developed a C-band LLRF control RFSoC platform with direct RF sampling, which targets Cool Copper Collider (\(C^3\)) and other C or S band LINAC research and development projects. In this paper, the architecture of the platform will be described. We have optimized the configuration of the data converter and characterized performance of them with RF pulses. The test results for some of the key performance parameters for the LLRF platform with our custom solid-state amplifier, such as phase and amplitude stability, will be discussed in this paper. 
\end{abstract}

\section{Introduction}
LLRF control circuits of LINACs typically convert the RF signals to base-band or vice versa using analogue mixers. The integrated analogue-to-digital (ADC) in Gen 3 RFSoC devices has a maximum input frequency of 6~GHz, which makes it possible to direct sample the C-band RF signal and perform the mixing digitally. The integrated DAC can generate RF signal up to 7 GHz with the integrated digital up mixer enabled. If the performance of the RFSoC data converters can meet the required, the full LLRF circuit can be implemented within an RFSoC. As the RFSoC devices has up to 16 ADCs and 16 DACs in a single device, so the control circuit of LINACs with large number of RF signals can be implement more compactly and cost-effectively. The  RFSoC data converters has demonstrated excellent RF performance in both first order Nyquist zone \cite{liu2021characterizing,liuRA} and higher order Nyquist zones \cite{liuRF} for other physics experiments. However, the LLRF control system of LINACs has unique requirements compared with other experiments. 

In this paper, the tests performed to determine the optimum data converter configurations will be summarized and the phase and amplitude stability characterization results will also be discussed based on the requirement of the LLRF control system of a C-band LINAC. 

The hardware platform used in this exercise is the ZCU216 evaluation board with an UltraScale+ XCZU49DR-2FFVF1760 RFSoC, which has 16 ADCs up to 2.5 Gigasample per second (GSPS) and 16 DACs up to 9.85 GSPS. With the device, a compact LLRF system can be realized with up 16 RF inputs up to 6 GHz and and 16 RF outputs up to 9.85 GSPS.

\section{RF Performance Evaluation}

Phase and amplitude stability is one of the most critical performance parameters for a LLRF system. In this experiment, the suitability of phase and amplitude is characterized in both continuous wave (CW) and pulsed modes.

\subsection{CW Performance Evaluation of Data Converters}

Loop-back test with different data converter configurations is the most efficient method to explore the optimum settings for a particular application. The most critical settings for the integrated data converters include the sampling speed for the data converts, digital mixer configuration on both up and down conversion sides and Nyqusit zone configurations. 

\begin{figure}[!htb]
   \centering
   \includegraphics*[width=1\columnwidth]{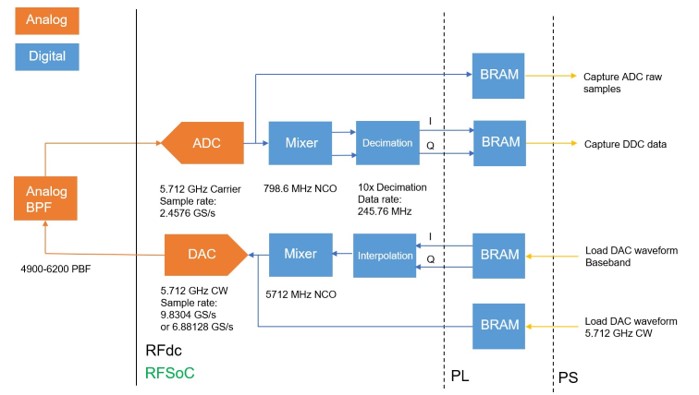}
   \caption{The loop-back test circuit for RFSoC DAC and ADC noise performance evaluation.}
   \label{fig:loopbacksetup}
\end{figure}

Figure \ref{fig:loopbacksetup} shows the circuit for loop-back evaluation. In this case, the evaluation has been performed with full up and down conversion data-paths on both ADC and DAC side. The sampling speed of ADC and DAC are 2.4576 GSPS and 5.89824 GSPS respectively. The sampling speeds are determined based on the structure of the clock tree and the requirement of the application. Custom base-band digital quadrature sequences can be loaded to dedicated block RAM (BRAM) in the FPGA of the RFSoC. The in-phase (I) and quadrature (Q) sequences are read by the interpolation block and then up-mixed at 5.712 GHz before converted by the DAC. The analogue RF signal generated
by the DAC is injected to the input of the integrated ADC via band-pass filter. The digitized signal is then down mixed at image of 5.712 GHz at the first Nyquist zone of the ADC. Then the signal is decimated and written to BRAM. The BRAM can be accessed from the software for further processing for performance evaluation. 

\begin{figure}[!htb]
   \centering
   \includegraphics*[width=1\columnwidth]{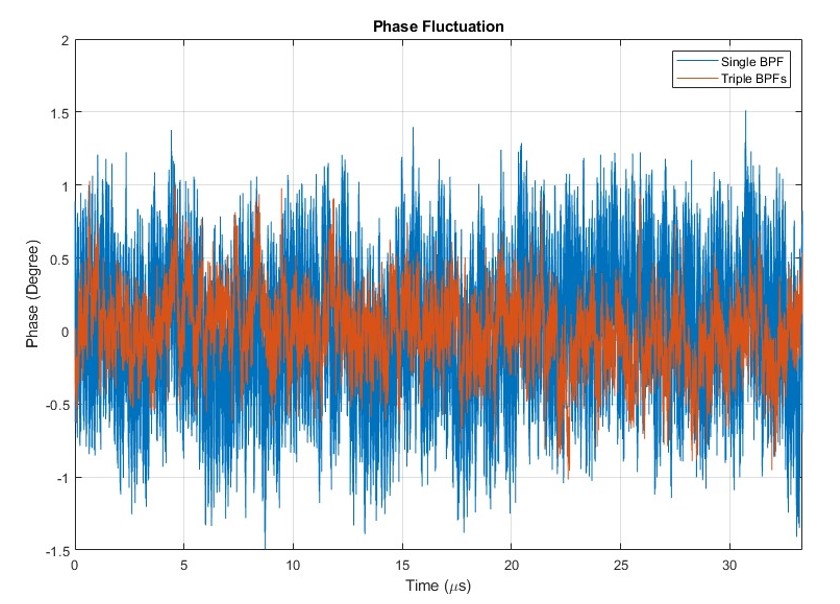}
   \caption{Phase fluctuation in 33 \(\mu\)s with 1 single BPF and 3 BPFs connected in series.}
   \label{fig:bpf_phase}
\end{figure}

\begin{figure}[!htb]
   \centering
   \includegraphics*[width=1\columnwidth]{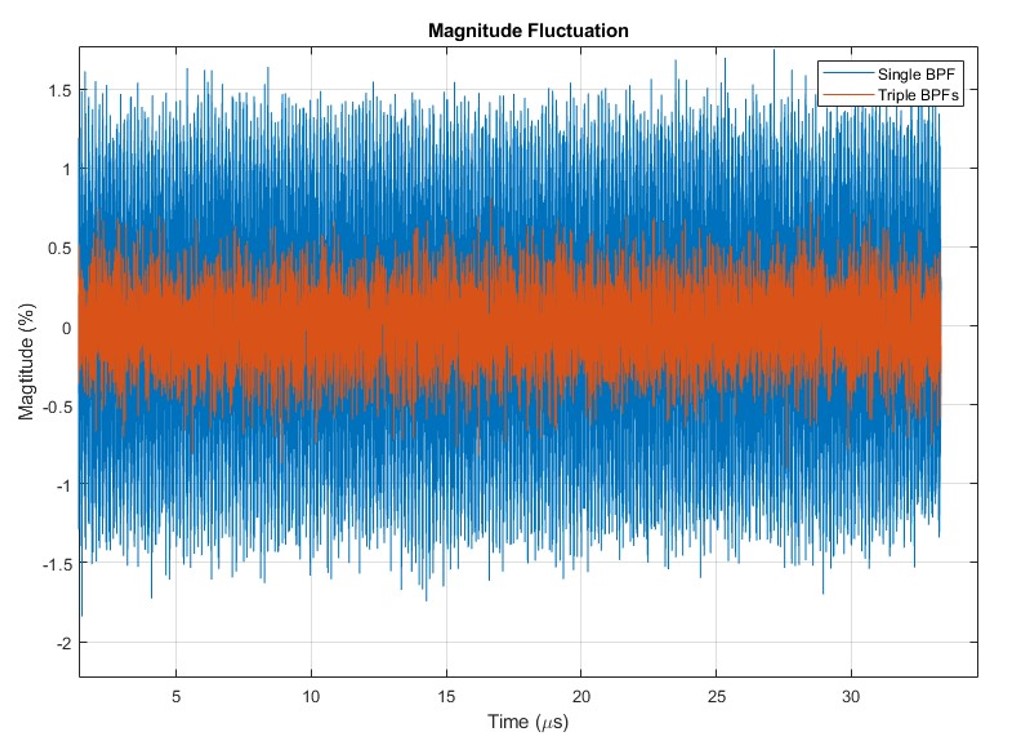}
   \caption{Amplitude fluctuation in 33 \(\mu\)s with 1 single BPF and 3 BPFs connected in series.}
   \label{fig:bpf_amp}
\end{figure}

The CW performance of the loop-back circuit has been performed by loading DC sequences to I and Q components the up-conversion side. The I and Q components are captured on the down-conversion side for 33 \(\mu\)s. Phase and amplitude of the down-converted signal are calculated by using the raw samples captured and shown in Figure \ref{fig:bpf_phase} and \ref{fig:bpf_amp}. To simplify the visualization of the fluctuations, the mean of the phase and amplitude has been subtracted from the raw values.

There are two traces in both phase and amplitude plots, one captured with only a single 4900 to 6200 MHz inline band-pass filter (BPF) and another captured with 3 of the filters connected in series. As figures show, the phase fluctuation with 3 BPFs is approximate 50\% less than with a single BPF and approximately 66\% less for the magnitude fluctuation. With 3 BPFs, the phase fluctuation is generally within \(\pm\) 0.5\textdegree and the amplitude fluctuation within \(\pm\) 0.5\%. The phase and amplitude stability levels are adequate for some of the C-band LINACs, but the performance can still be improved by box filter or compensated by other techniques. This test shows that the BPF filter has a significant impact in the stability performance. The VNA measurement shows that the side-band rejection of 3 BPFs combined is approximately 40 dB higher than the single BPF. 

As the RF frequency is located in the second order Nyquist zone of the DAC and fifth order Nyquist zone of the ADC, the harmonics and other spurs can fold back to the first Nyqusit zone and result in higher noise level. Therefore, custom BPFs with high side-band rejection should be applied for the optimum stability performance.  

\subsection{Pulse Top Flatness Evaluation}

The klystron of the LINAC is generally driven by the pulsed RF output of LLRF circuit after amplified by solid state amplifier (SSA). Some of the LINACs has specifications for the flatness of the RF pulse generated by the SSA. In this exercise, the flatness of the RF pulse generated by the DAC and the high power RF pulse after attenuation will be evaluated. 

\begin{figure}[!htb]
   \centering
   \includegraphics*[width=1\columnwidth]{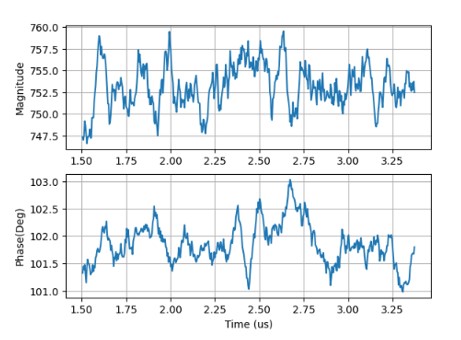}
   \caption{Amplitude and phase fluctuation on pulse top with DAC to ADC loop-back circuit.}
   \label{fig:pt_lb}
\end{figure}

The test circuit in Figure \ref{fig:loopbacksetup} with 3 BPFs has been used, but the rectangular pulse with 2 \(\mu\)s duration is loaded to DAC data-path at 60 Hz. The I and Q components on the ADC side are captured and converted to amplitude and phase. The amplitude and phase fluctuations on top of pulse are shown in Figure \ref{fig:pt_lb}. The amplitude values are the direct calculations of the digital data and the values are is on the lower side of the full range of ADC due to the limits on circuit configuration. The amplitude fluctuation measured by the ratio of the standard deviation and the average value is as low as 0.34 \%. The standard deviation of phase fluctuation is approximately 0.37\textdegree. The fluctuation performance can still be improved by using more range of data converters. 

\begin{figure}[!htb]
   \centering
   \includegraphics*[width=1\columnwidth]{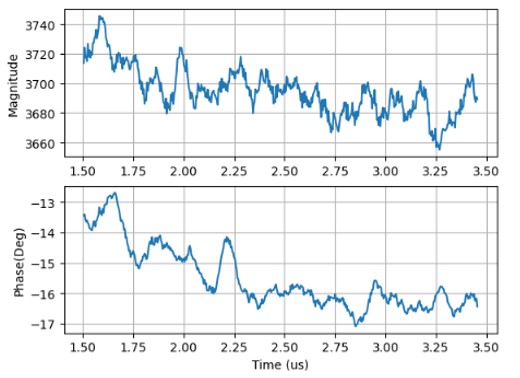}
   \caption{Amplitude and phase fluctuation on pulse top with SSA and attenuation and coupler.}
   \label{fig:pt_ssa}
\end{figure}

The test has also been performed with the SSA, which has approximately 300 W output power. The SSA amplifies the output of the DAC and then the signal is attenuated to the level within the input range of ADC. The amplitude and phase fluctuations are shown in Figure \ref{fig:pt_ssa}. The ratio of the standard deviation and the average value of the amplitude is 0.44 \%, which is marginally higher than the test without the SSA. The standard deviation of phase values is as high as 1.03\textdegree. As the phase plot shown in Figure \ref{fig:pt_ssa}, there is a phase drift on top of the fluctuation introduced by the SSA and that results in the significant high phase fluctuation measured in this case. For applications with more stringent pulse top flatness requirement, the the LLRF needs to be designed to have the capability of compensating the drift. 

\subsection{Pulse-to-pulse Fluctuation Evaluation}

\begin{figure}[!htb]
   \centering
   \includegraphics*[width=1\columnwidth]{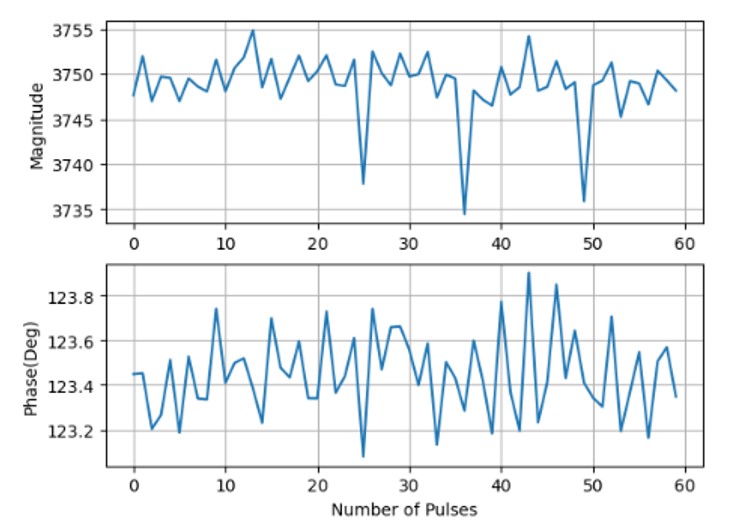}
   \caption{Average amplitude and phase fluctuation of 60 consecutive RF pulses in 1s. }
   \label{fig:pp_ssa}
\end{figure}

Pulse-to-pulse fluctuation is the most critical parameter for inter pulse feedback LLRF system. In this exercise, 60 consecutive RF pulses are captured with the LLRF system with SSA. The average values for amplitude and phase are calculated from the sample captured in I and Q format and shown in Figure \ref{fig:pp_ssa}. The amplitude fluctuation is approximately 0.09\% and the phase fluctuation is around 0.18\textdegree, which is equivalent to 87.54 femtoseconds. The fluctuation is lower than the 0.1\% and 0.3\textdegree  RF phase requirement of the Cool Copper Collider (\(C^3\)) \cite{emilio} and many other LINACs. 

\subsection{Pulse-to-pulse Fluctuation Evaluation in S-band}

\begin{figure}[!htb]
   \centering
   \includegraphics*[width=1\columnwidth]{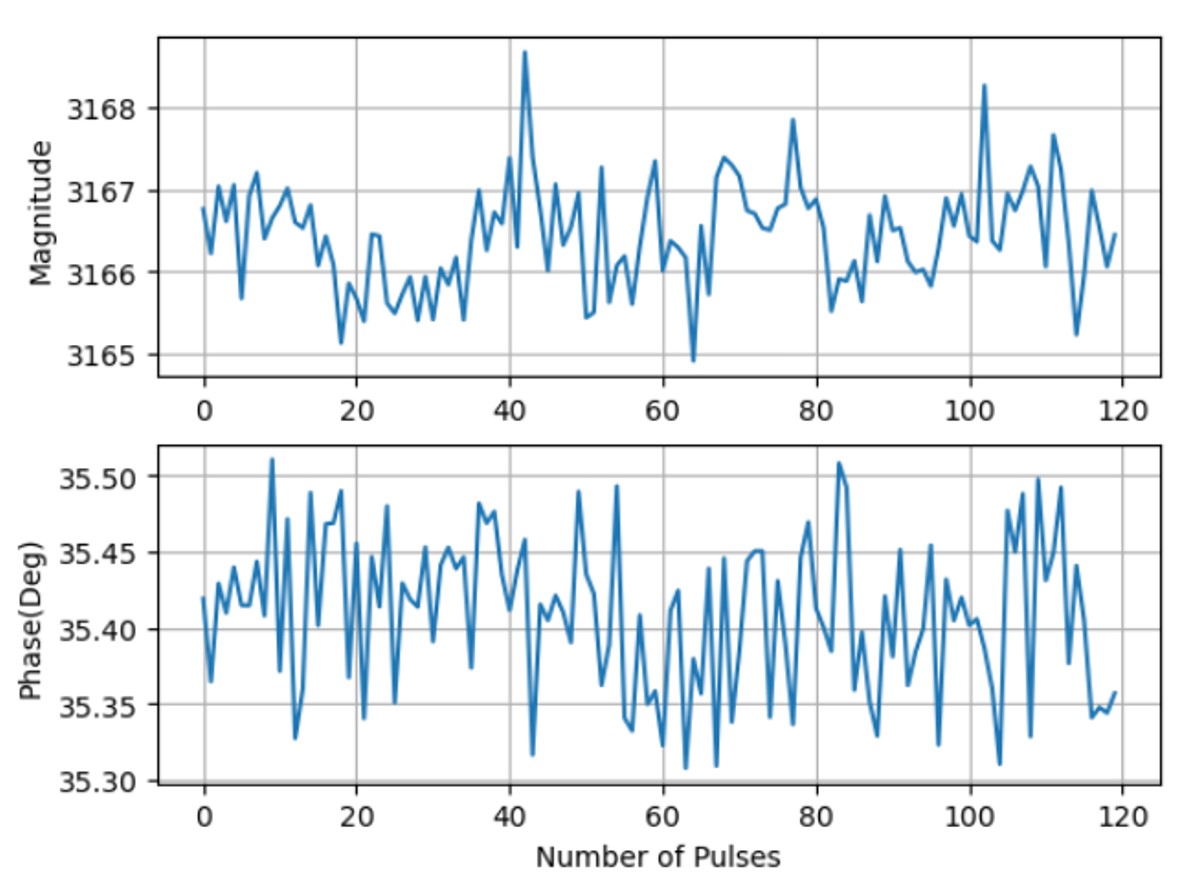}
   \caption{Average amplitude and phase fluctuation of 120 consecutive RF pulses in 1s. }
   \label{fig:slb}
\end{figure}

Similar inter pulse fluctuation has been performed with S-band application with RF frequency centred at 2.856 GHz. In this case, the test has been performed with ZCU208 evaluation board using a different GEN 3 RFSoC device with DAC and ADC sampling at 5.89824 GSPS and 4.9152 GSPS respectively. The loop-back circuit with a BPF centred at 2.856 GHz has been used. The pulse rate has been set to 120 Hz, which is the pulse rate of LCLS. As Figure \ref{fig:slb} shows, amplitude and phase fluctuations in 1 s are 0.02\% and 0.05\textdegree (equivalent to 48.6 femotoseconds).

\section{Summary}

 The exercise has demonstrated that the LLRF system with the custom designed SSA can deliver a phase jitter as low as 87.54 femtoseconds, which is considerably better than the requirement of \(C^3\).The conventional LLRF circuit with discrete data converters and RF mixer can offer higher phase stability, but the high integration level of RFSoC based system offers a significantly more compactly and cost-effectively solution. The system is highly flexible in tuning RF frequency, changing the pulse shape and implementing custom control schemes in firmware and software.  This system has been fully implemented with commercial hardware and the phase noise could be improved by using clock circuity optimized for a particular range and better filtering circuit when developing a custom RFSoC based system. 
%
%
\ifboolexpr{bool{jacowbiblatex}}%
	{\printbibliography}%
	{%
	
	
} 
%
%


\end{document}